\documentclass[11pt]{revtex4}
\usepackage{amsfonts}
\usepackage{amssymb}
\usepackage{amsfonts}
\usepackage{amsmath}
\usepackage{amssymb}
\usepackage{amsthm}
\usepackage{epsf}
\newcommand{\Hh}{\displaystyle\frac12}

\def\Smm{^3S_1^{--}}
\def\Dmm{^3D_1^{--}}

\def\Spp{^3S_1^{++}}
\def\Dpp{^3D_1^{++}}
\def\Smm{^3S_1^{--}}
\def\Dmm{^3D_1^{--}}
\def\Pte{^3P_1^{+-}}
\def\Pto{^3P_1^{-+}}

\def\Pse{^1P_1^{+-}}
\def\Pso{^1P_1^{-+}}

\def\mbf #1{{\mbox{\boldmath$#1$}}}
\newcommand{\bqn}{\begin{eqnarray}}
\newcommand{\eqn}{\end{eqnarray}}
     \font\tenbifull=cmmib10 scaled 1200 
     \font\tenbimed=cmmib9
     \font\tenbismall=cmmib7
       \def\bmit{\fam9 }
\mathchardef\bbkappa="7114
\mathchardef\bbgamma="710D
\mathchardef\bbrho="711A
\mathchardef\bbsigma="711B
\mathchardef\bbtau="711C
\mathchardef\bbvarrho="7125
\mathchardef\bbvarsigma="7126
\mathchardef\bbxi="7118
\def\boldgamma{{\bmit\bbgamma}}

\def\CG {{\cal G}}

\usepackage[dvips]{graphicx}
\def\bk{{\mbox{\boldmath$k$}}}

\def\bp{{\mbox{\boldmath$p$}}}

\begin{document}
\title{Solving   the  Bethe-Salpeter Equation  in
Euclidean Space\thanks{Based on materials of the contribution
"Relativistic Description of Two- and Three-Body Systems in Nuclear Physics",
 ECT*, October 19-13, 2009}}
\author {S.M. Dorkin}\affiliation{International University Dubna, Dubna, Russia, \\
 Bogoliubov Lab. Theor.
Phys. JINR, 141980  Dubna, Moscow reg., Russia}
\author{L.P. Kaptari\footnote{Supported
through the program Rientro dei Cervelli of the Italian Ministry
of University and Research}}\affiliation{Bogoliubov Lab. Theor.
Phys. JINR, 141980  Dubna, Moscow reg., Russia \\Dept. of Phys., Univ. of Perugia and
      INFN, Sez. di Perugia, via A. Pascoli, I-06123, Italy }
\author{C. Ciofi degli Atti}\affiliation{Dept. of Phys., Univ. of Perugia and
      INFN, Sez. di Perugia, via A. Pascoli, I-06123, Italy}
\author{B. K\"ampfer}\affiliation{Forschungszentrum Dresden-Rossendorf, PF 510119, 01314 Dresden, Germany}


\begin{abstract}
 Different approaches to solve the spinor-spinor  Bethe-Salpeter (BS) equation
 in   Euclidean space are considered.
 It is argued  that the complete set of Dirac matrices is the most appropriate
 basis to define the partial amplitudes and to solve numerically the resulting
  system of equations with realistic
 interaction kernels. Other representations can be obtained by  performing  proper
 unitary transformations. A generalization of the iteration method for  finding
 the energy spectrum of the BS equation is discussed and examples of concrete calculations
 are presented. Comparison of relativistic calculations with available experimental data
  and with
 corresponding non relativistic results together with an analysis of the role of
  Lorentz boost effects and relativistic corrections are presented.

A novel method related
to the use of hyperspherical harmonics is considered for
a   representation of the  vertex functions suitable  for numerical calculations.

\end{abstract}

\maketitle

\section{Introduction}

\label{intro}
The interpretation of many modern experiments
requires often a covariant description of   two-body systems. This is
 due either to  the high precision that calls for an inclusion of all possible
corrections to a standard (possibly non relativistic) approach, or due
to the high energies and momenta involved in the investigated processes.
In the subatomic field the most obvious
examples are the properties and structure of the deuteron
and, to some extent, the mesons, if the these are  treated as   quark antiquark systems.
Also the charm degree of freedom is a recent topic of analysis in heavy ion
experiments. The CBM experiment of the future FAIR project at GSI
will investigate highly compressed dense matter in nuclear collisions with a
beam energy range between 10 and 40 GeV/u. An important part of the
hadron physics project is devoted to extend
the SIS/GSI program for the search of  in-medium modifications
 of hadrons to the heavy quark
sector providing a first
insight into charm-nucleus interaction. Thus, the possible modifications
of the properties of open and hidden charm mesons in a hot and dense
environments is matter of recent studies.

Within a local quantum field theory the starting point of a relativistic
covariant description of  bound states of two particles is the
Bethe-Salpeter (BS)  equation. Due to known difficulties related to
the analytical properties of the BS amplitude in Minkowski space,
the procedure of solving the  equation
 turns out to be a rather  involved task. Up to now, the BS equation
including realistic
  interaction kernels has been solved either in Euclidean space
  within the ladder approximation
  (see e.g. Refs.~\cite{rupp}-\cite{maris}  
  and references  quoted therein)
 or adopting additional approximations
 of the BS equation itself~\cite{gross}-\cite{karmanov_PR}.
 The
procedure of solving  numerically the BS equation  has been revisited  in Ref.~\cite{fb-tjon},
and a reduction of the BS equation to an equation of
the Light Front form has been
 proposed in Ref.~\cite{tobias}. Moreover, a detailed investigation of
the  solution and the properties of  two spinor
particle in a  bound state within  Light Front Dynamics  have been reported
in Refs.~\cite{karmanov,gsalme}.

The present paper represents a brief review of mathematical and numerical
procedures  of solving the BS equation in Euclidian space.  Advantages and disadvantages  of
different methods depending of the  feature of the attacked problem are discussed.
Section \ref{chap2} surveys basic features in solving the BS equation.

In Subsection \ref{diracsec}, it is demonstrated that the
Dirac representation of the BS amplitude with subsequent use of the iteration method for solving the
system of integral equations for the partial amplitudes, is the simplest and  most appropriate way to obtain
a numerical solution. However, the physical interpretation of results of calculations
within such a representation is not so transparent. The spin angular harmonics basis,
as discussed in Subsection \ref{spinang},
 provides a more transparent interpretation of the partial amplitudes in terms of familiar notations
 used in  non relativistic approaches. Obviously, in practice, one can combine the
  two representations; the numerical solution
may be obtained within the Dirac representation and then, by a unitary transformation,
transferred to the spin angular basis.
A covariant form, i.e. a frame independent  representation of the BS equation is
 presented in Subsection \ref{secov}.
The possibility to obtain
 approximate analytical expressions  of  matrix elements within the
BS formalism is discussed in Subsection \ref{oneiter} in the context of the
one-iteration approximation procedure.
 A generalization of the iteration procedure
 to find the energy spectrum of the BS equation is presented in Section \ref{exhaus}.
 The use of the hyper spherical
harmonics  basis to interpret the BS amplitude in form of
one-dimensional numerical arrays, suitable for solving the BS and Dyson-Schwinger equations for
  quark-antiquark mesons, is
discussed in Subsection \ref{hyp}.

\section{Solving the BS equation}
\label{chap2} \noindent

 The BS amplitude  for two spinor particles  $A$ and $B$ is defined as \cite{BS}
\begin{eqnarray}
\Phi(x_1,x_2)_{\alpha\beta} = \langle 0|
T{\psi}^A_\alpha(x_1){\psi}^B_\beta(x_2)|AB\rangle\equiv {\rm
e}^{-i P \cdot X}\Phi(x)_{\alpha\beta}, \label{BSA}
\end{eqnarray}
where $|AB\rangle$ denotes the bound or scattering state
of the $AB$ system, ${\psi}^A(x_1)$ and ${\psi}^B(x_2)$ are the fermion field operators
in the Heisenberg representation with spinor indices
 $\alpha, \beta$, and $x=x_1-x_2$ is the relative coordinate.
 The amplitude (\ref{BSA}) obeys the BS equation which in the ladder approximations
 reads
\begin{eqnarray}
\Psi(p) = i \sum\limits_{b}g^2_b \int\frac{d^4p'}{(2\pi)^4}
\frac{S(p_1)\gamma_b(1)\Psi_D(p')\gamma_b(2){\tilde S}(p_2)}
{(p-p')^2-\mu_b^2+i\varepsilon}, \label{eqrl}
\end{eqnarray}
 where  $b$ denotes the type of exchanged mesons
 (e.g. $\pi$, $\rho$, $\omega$, $\sigma$, $\delta$ and  $\eta$ for the deuteron problem);
 $\gamma_b$ are the corresponding  nucleon meson vertices; the nucleon momenta are
 $p_{1,2}=P/2 \pm p,~P$, and $S(p)$ is the free nucleon propagator.
 The amplitude (\ref{BSA}) is redefined as $\Psi=-\Phi\,
 U_C$, where $U_C$ is a charge conjugation-like matrix
 ($U_C=i\gamma_2\gamma_0$, or $ U_C=\gamma_1\gamma_3$)
   ${\tilde S}(q) = U_C\, S^T(q)\, U_C$.

 From Eqs. (\ref{BSA}) and (\ref{eqrl}) it can be seen
 that in Minkowski space the BS amplitude is a rather complicate object, being  a
 $4\times4 $ matrix in the spin space and having
  cuts \cite{wick} and poles along the real axis of the relative energy in
  configuration space.
 Usually, instead of solving directly the matrix equation (\ref{eqrl}),
 one decomposes the BS amplitude $\Psi$
 over a complete set of basis matrices
  and solves the equation for the coefficient functions  of such a decomposition.
   Generally, these   coefficient functions form a system of four-dimensional
   integral equations.
  In order to reduce their dimension  the coefficient functions
  are also decomposed  over a complete set of orthogonal  functions in configuration space.
  Such a procedure is mathematically correct if the original amplitude is
 an analytical function of its arguments, which  is not the case in Minkowski
 space. To obtain a  completely regular equation one performs the Wick rotation~\cite{wick}
 to Euclidian space  where the unknown quantities (i.e. the partial BS amplitudes)
 are defined  along the imaginary axis of the relative energy.
 The Mandelstam technique \cite{mandelstam} relates the matrix elements
  of observables in physical processes
 with the BS amplitudes in Euclidian space.
However, often, the BS amplitudes are needed directly
in Minkowski space, so that  an algorithm for
 analytical continuation of the Euclidian solution in
the whole plane of the relative
energy $p_0$ or a numerical procedure of  inverse Wick rotation
back to Minkowski space are demanded. These procedures  are not yet well established and
 are still  under consideration by many theoretical
 groups\footnote{
 In some simple cases it is possible to solve the BS equation directly in
 Minkowski space by using the
 Nakanishi representation \cite{nakan} and projecting the BS equation
 on to the light front coordinates \cite{karmanov,karmanov_PR}.}.
 In the present paper, we focus our attention on mathematical and numerical methods
 of solving the BS equation in Euclidian space with
 realistic interaction kernels in the ladder approximation.
 The problem of finding the
 solution in Minkowski space is not considered here.

\subsection{Dirac representation}\label{diracsec}
The most appropriate choice of the basis  for
the representation of the amplitude $\Psi(p)$ in  spinor space  is the complete set
of Dirac matrices $\gamma_i$ ($i=1\ldots 16$), which contains
a minimal number of $4\!\times\!4$ matrices and, consequently,
significantly  simplifies calculations of corresponding matrix elements and
traces. In this representation the amplitude becomes
\begin{eqnarray}
  \Psi(p) = \hat{  1} \psi_s(p) + \gamma_5 \psi_p(p) + \gamma_{\mu}
\psi_v^{\mu}(p) + \gamma_5 \gamma_{\mu} \psi_a^{\mu}(p) +
\sigma_{\mu\nu} \psi_t^{\mu\nu}(p), \label{amp}
\end{eqnarray}
where the subscripts of the coefficient functions $\psi_i$ refer to
the  transformation properties,
i.e., these coefficients can be scalar (s), pseudo-scalar (ps),
vector (v), axial-vector (a)  and
tensor (t) functions, respectively.
Further, in order to eliminate the angular dependence,
the coefficients $\psi_i$  are  decomposed correspondingly over the
spherical $\rm Y_{LM}$ and vector angular $\rm {\bf Y_{1M}^L}$ harmonics.
 For instance, in the $^3S_1-^3D_1$ channel (that is for the deuteron case)
 the amplitude reads
\begin{eqnarray}
&&\Psi_D(p) = \frac{1}{|{\bf p}|} \left \{ \gamma_5\,
\psi_{p1}{\mbox{\large {Y}}}_{1M} + \gamma_5 \gamma_0\,\psi_{a1}^0
{\mbox{\large {Y}}}_{1M} -\psi_{v1}(\boldgamma, {\mbox{\large \bf
{Y}}}_{1M}^{1}) -\psi_{a0}\gamma_5(\boldgamma,
{\mbox{\large \bf {Y}}}_{1M}^{0})\right.\nonumber \\
&&\left.-\psi_{a2}\gamma_5(\boldgamma, {\mbox{\large \bf
{Y}}}_{1M}^{2}) -2i\psi_{t1}^0\gamma_0(\boldgamma, {\mbox{\large \bf
{Y}}}_{1M}^{1}) +i\psi_{t0}(\boldgamma,[\boldgamma,{\mbox{\large \bf
{Y}}}_{1M}^{0}]) +i\psi_{t2}(\boldgamma,[\boldgamma,{\mbox{\large \bf
{Y}}}_{1M}^{2}]) \right \}. \label{amp6}
\end{eqnarray}

By placing eq. (\ref{amp6}) in to eq. (\ref{eqrl}), calculating
the corresponding traces and carrying out angular integrations
one obtains a system of eight two-dimensional integral equations of the form
\begin{eqnarray}
\psi_{p1}  (p_0, {\bf |p|})&= &
\hat {\sf K}_1\left \{ \left
[ m_N^2 +p^2 - \frac{1}{4}M^2_d \right ] \psi_{p1}
  - 2m_Np_0\psi_{a1}^0 + \right.  \nonumber\\
 &+& \left .2m_N{|\bf p|}\left (\frac{1}{\sqrt{3}}\psi_{a0}
-\sqrt{\frac{2}{3}}\psi_{a1}   \right )
+2M_d|{\bf p}|\left (\frac{1}{\sqrt{3}}\psi_{t0}
-\sqrt{\frac{2}{3}}\psi_{t2}  \right ) \right \},\label{b1}
\\
\psi_{a1}^0 (p_0, {\bf |p|})
&= &\hat {\sf K}_1\left \{ -2m_Np_0\psi_{p1}
+\left [ m_N^2 - p^2 + 2 p_0^2 - \frac{1}{4}M^2_d \right ]
 \psi_{a1}^0 -\right .
  \\   \hspace*{2cm}
  &-& \left . 2p_0|{\bf p}|\left (
\frac{1}{\sqrt{3}}\psi_{a0} -\sqrt{\frac{2}{3}}\psi_{a2} \right )\right \},
\label{b2}\\
\psi_{v1}  (p_0, {\bf |p|}) &=&\ \hat {\sf K}_1\left \{ \phantom{\frac12}\ldots\ldots\right \}.
\label{bb2}
\end{eqnarray}
where  $\hat {\sf K}_1$ is an integral operator whose explicit expression
depends   on the type of the exchanged meson. For instance, for   scalar mesons it reads
\begin{eqnarray}
\hat {\sf K}_1 \psi_L (p_0,|{\bf p}|) =
-i g^2_\sigma D(p_0,|{\bf p}|)
\int\!\frac{dp_0^\prime d|{\bf p}^\prime||{\bf p}^\prime|^2}{(2\pi)^3}
Q_L(y)\psi_L(p_0^\prime,|{\bf p}^\prime|),
\label{oper2}
\end{eqnarray}
and
\begin{equation}
D(p_0,{\bf p})=\displaystyle\frac{1}{(E_{\bf p}^2-p_0^2-\frac14
 M_d^{ 2})^2-p_0^2  M_d^{ 2}}
\label{scprop}
\end{equation}
is the scalar part of two propagators.
In Eq. (\ref{oper2}), $Q_L(y)$ are the Legendre functions with the
argument $
y = \frac{ |{\bf p}|^2 + _{}|{\bf p}^\prime|^2 + \mu^2_B -
(p_0-p_0^\prime)^2 } {2|{\bf p}||{\bf p}^\prime|}
$ with $\mu_B$ as the mass of the exchanged meson.
Analogous expressions can be obtained for exchanges of pseudo-scalar and vector
  mesons.

The resulting system of integral equations (\ref{b1})-(\ref{bb2}), after performing the
Wick rotation, is solved by using the iteration procedure. Note  that, since
 the eigenvalue $M_d$  (i.e. the deuteron mass) enters nonlinearly in
 Eqs. in (\ref{b1})-(\ref{bb2}) (cf. Eq. (\ref{scprop})),
 the iteration procedure is highly prevented.
  Usually, one fixes the binding energy at the experimentally known  value
  and searches the solution relative to the coupling constants $g$ which enter linearly in
   Eqs. (\ref{b1})-(\ref{bb2}) and for which the iteration procedure can be employed.

    We solved by this method
   the BS equation for the deuteron with a realistic kernel   within the
one-boson exchange interaction  with six exchange mesons \cite{umnikov}.
The numerical solution consists of eight partial amplitudes in the  form of
two-dimensional arrays indexed by  $|{\bf p}|$ and $p_0$,
 which can be further   used to analyze the
 properties of the deuteron \cite{static} and to compute matrix elements of different
 processes \cite{static}-\cite{fsiHe3}. It should be noted  that,
 besides advantages  (simple Clifford algebra with Dirac matrices,
 fast convergence of the  iteration procedure) within the Dirac representation,
the physical interpretation of the partial amplitudes (\ref{amp})-(\ref{amp6}) and
matrix elements of observables   in terms of familiar nonrelativistic
quantities, Lorentz boost effects  and comparisons with other approaches
(light front dynamics \cite{karmanov,karmanov_PR}, Gross equation \cite{gross} etc.)
are difficult.

\subsection{Spin angular basis}\label{spinang}
\begin{figure}
  \includegraphics[width=0.41 \textwidth,height=0.3\textheight]{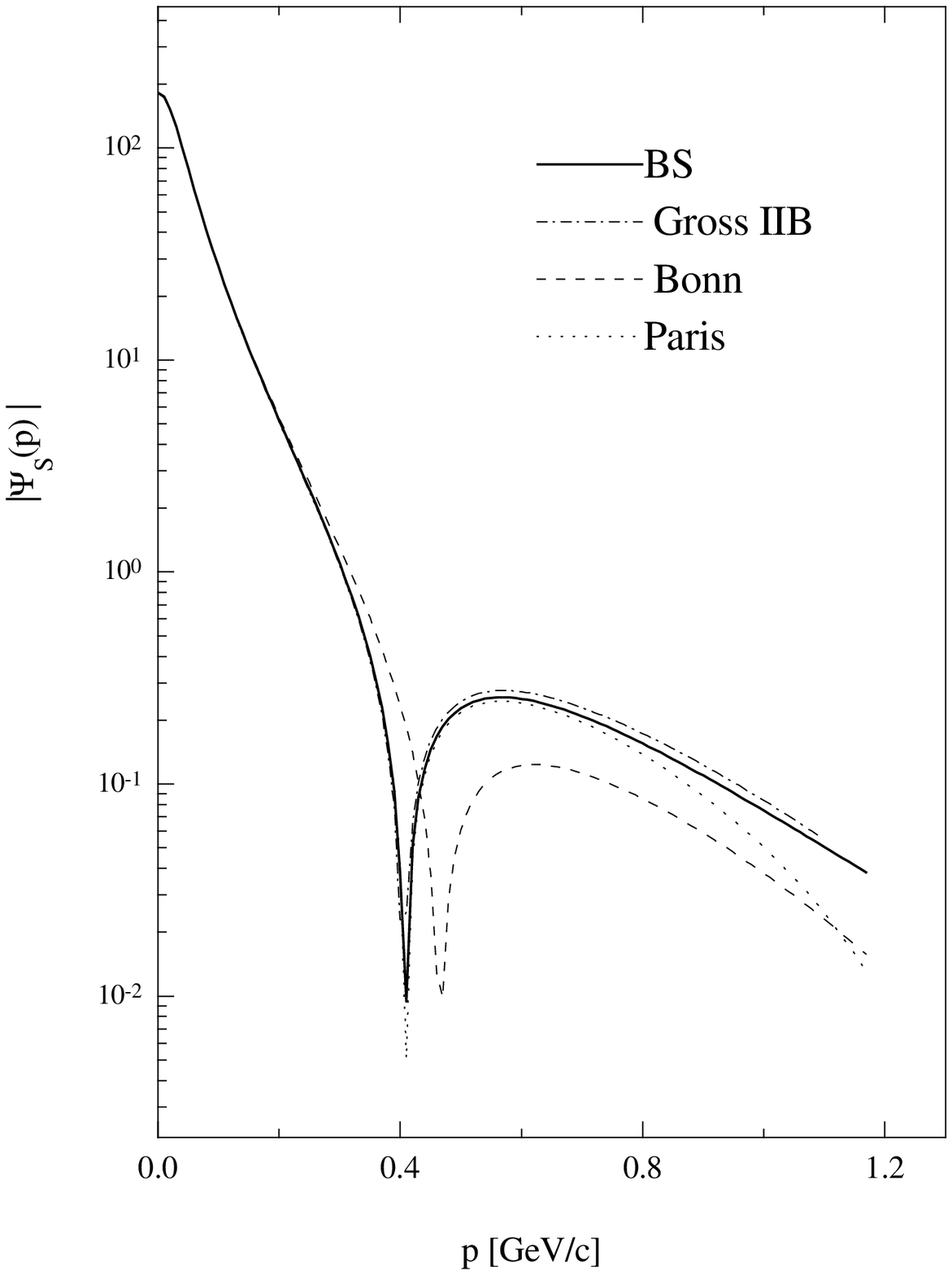}
  \hfill\includegraphics[width=0.41\textwidth,height=0.3\textheight]{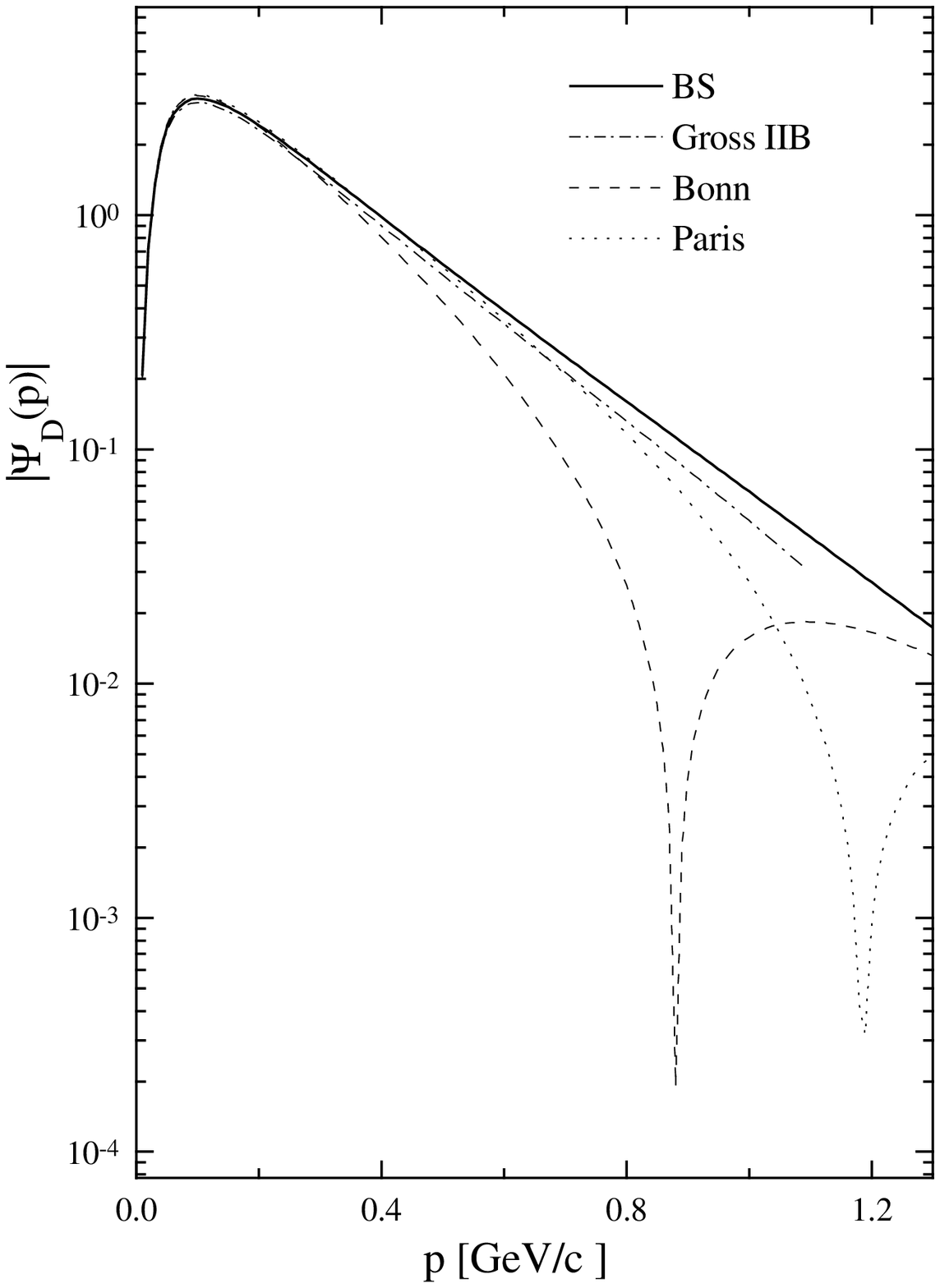}
\caption{Comparison of the deuteron  $S$ (left panel) and $D$-waves (right panel)
obtained within different approaches: Bethe Salpeter \cite{umnikov} (solid lines),
Gross equation \cite{gross} (dot-dahes lines), non relativistic Schr\"odinger
approach with Bonn (dashed lines) and Paris (dotted lines) potentials.} \label{waves}
\end{figure}

The basis of spin-angular harmonics
${\Gamma}_M^{\alpha}(\mbf{p})$\cite{rupp,tjon,kubis} allows  for a more transparent
interpretation of the solution. It is constructed from the complete set of solution of
Dirac equation for free nucleons
\bqn
{\Gamma}_M^{\alpha}(\mbf{p}) U_C
= (-)^{\rho_1+\rho_2} \ i^L \sum \limits_{\mu_1 \mu_2 m_L}\; (L
m_L S m_S | J M) \; (\Hh \mu_1 \Hh \mu_2 | S m_S)
 Y_{L m_L}({\hat{\mbf{p}}})\;
{U_{\mu_1}^{\rho_1}}(\mbf{p})\;
{U_{\mu_2}^{\rho_2}}^{T}(-\mbf{p}), \label{gammadecomp} \eqn
where ${\hat {\mbf{p}}} = {\mbf{p}}/{|\mbf{p}|},$ and the spinors
$U_{\mu_i}^{\rho_i}(\mbf{p}_i)$  are the solutions of the free Dirac equation
for the particle "i" with spin projection
$\mu$  and positive or negative   $\rho$-spin \cite{kubis};
$\alpha$ denotes other quantum numbers of the system; $\alpha=\{LSJ\rho_1
\rho_2\}$, where $L$ is the relative orbital momentum, $S$ the total spin and
$J$ the total momentum of the system. Often, for  $\alpha$ the spectroscopic
notation   $\alpha\equiv~^{2S+1}L^{\rho_1
\rho_2}_{J}$ is employed \cite{kubis}. The explicit expressions for the spin-angular
harmonics can be found, e.g., in Ref. \cite{static}. Evidently, the Dirac and the
spin-angular harmonics basis are connected via a  unitary transformation, presented
explicitly in Ref. \cite{static}.
Now the strategy can be formulated as follows: the BS solution and matrix
elements of   observables  are  obtained within the
Dirac representation, then by using the unitary transformation to the
spin-angular basis, results are  rewritten in a more familiar form in
terms of spectroscopical partial amplitudes. This  provides a
 more clear physical meaning  of the obtained expressions and allows for a
detailed comparison with
other relativistic and non relativistic approaches. Moreover, with partial BS amplitude
within the spectroscopical classification one can
"a priori" estimate the contributions of different terms. So, those amplitudes
which have a direct analogue in the non relativistic approach shall provide the
main contribution at low energies. Other  partial waves can contribute only at higher
energies and higher momentum transfer.
For instance, for the deuteron the eight components can be classified in   three groups:
1) the main components with two positive values of the $\rho$ spin,
$\phi_{\,\Spp}$, $\phi_{\,\Dpp}$, which are the relativistic generalization
of then familiar $S$ and $D$-waves of the deuteron
 2) four "negative" $P$-waves, with one  positive and another  negative $\rho$-spin:
$\phi_{\,\Pse}$,$\phi_{\,\Pso}$, $\phi_{\,\Pte}$,
$\phi_{\,\Pto}$, and 3) two negligible small components with both
$\rho$-spins negative $\phi_{\,\Smm}$, $\phi_{\,\Dmm}$\cite{static}.
Usually,  the later ones are disregarded in most   calculations.
As an example of the obtained solution
  we present in Fig.\ref{waves} the positive components
$S^{++}$ and $D^{++}$ of the BS vertex (solid lines) and compare them  with the non relativistic
results obtained from Schr\"odinger equation with Bonn and Paris potentials (dashed
and dotted lines, respectively) and with results provided by the relativistic three-dimensional
Gross equation \cite{gross}.  As expected, at low values of the relative momenta $p$
all approaches provide basically the same results. Differences occur  at large values of
 $p$, where the validity of non relativistic approaches becomes questionable.
 With this solution we have analyzed different reactions  with the deuteron in electromagnetic
 \cite{umnikov,fsiHe3,yscaling}, weak \cite{solar}  and hadronic processes \cite{dp,our_chex,our_brkp} with
 the result that calculations within the BS formalism provide a much better
 agreement with experimental data in comparison with non relativistic approaches.

\subsection{Covariant representation}\label{secov}
\begin{figure}
  \includegraphics[width=0.39\textwidth,height=0.29\textheight]{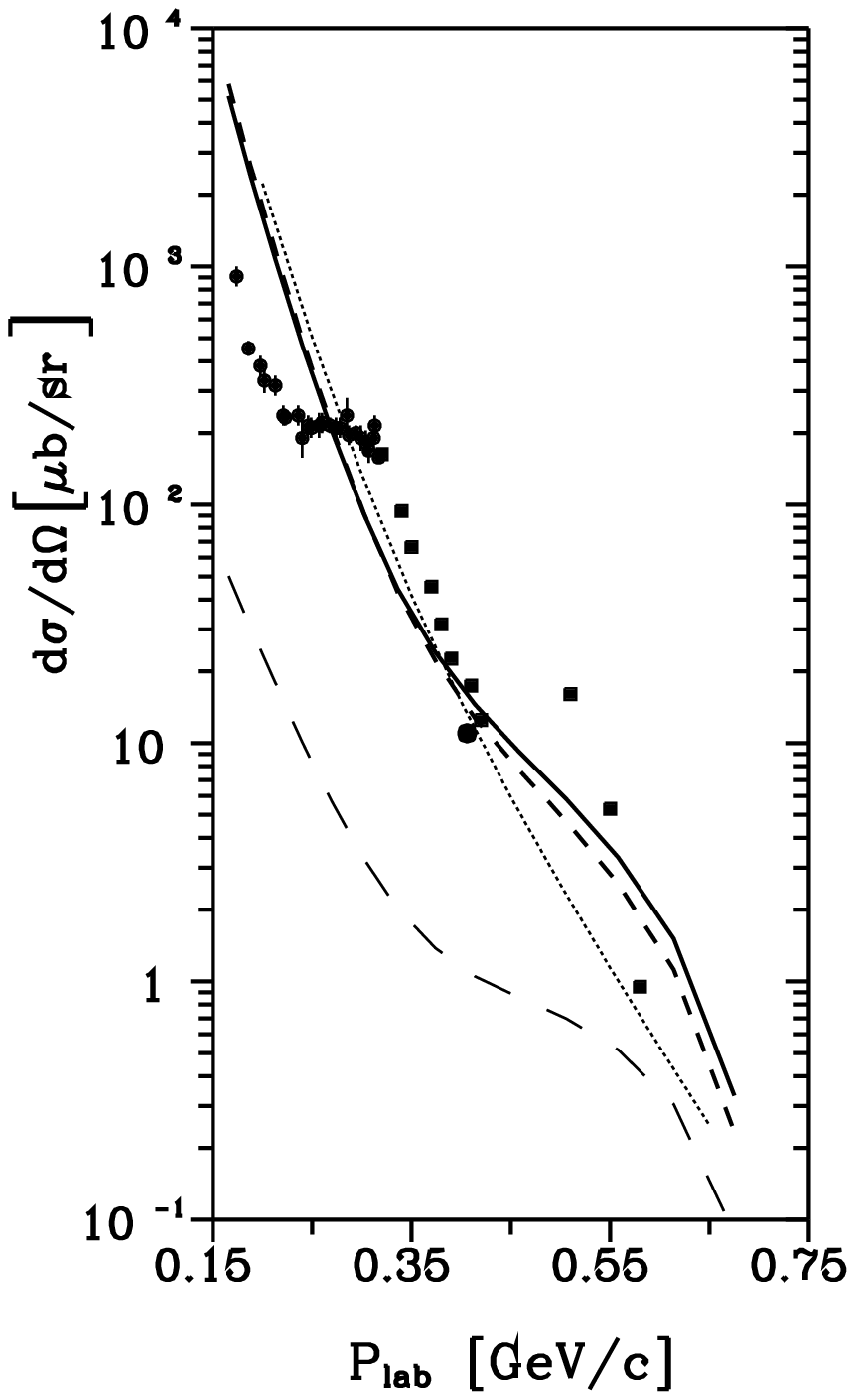}
  \hfill\includegraphics[width=0.39\textwidth,height=0.29\textheight]{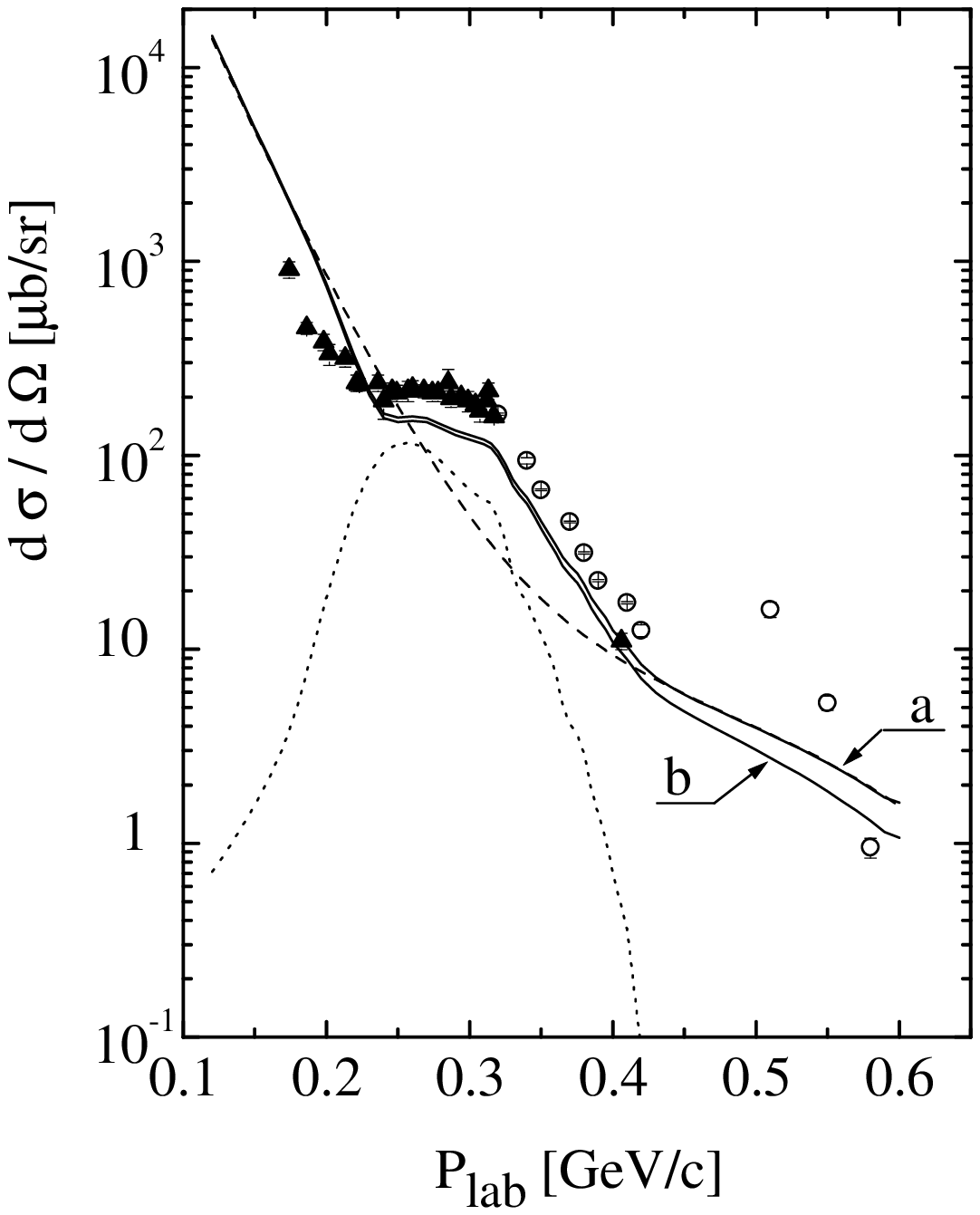}
\caption{ The spin averaged cross section $d\sigma/d\Omega$ for the backward elastic $dp$
 scattering in the center of mass as a function of the momentum of the detected
proton in the laboratory system. Left panel:
Dashed line - contribution of the positive BS waves; long-dashed line -
 Lorentz boost effects, solid line - full BS calculations;
 dotted line-results of  the non relativistic approach with the Bonn
potential wave function. Experimental data are from \cite{exppD}.
Right panel: dashed line - contribution of the one-nucleon exchange mechanism only;
 dotted line - contribution of the triangle diagram ($\Delta$-isobars);
 solid lines - the full BS calculations with $P$ waves (a) and
 without $P$  waves (b). } \label{pD}
\end{figure}

 It should be noted that the non relativistic Schr\"odinger
 approach belongs to the so-called "instant-form" of dynamics~\cite{diracForms}, i.e. the
 solution of the Schr\"odinger equation  is obtained in the center of mass system
 of the two-body system; being invariant only relative to the Lorentz transformation which
 leave the plane $t=const$ unchanged, it can not be boosted to another system of reference.
 This means that in  non relativistic calculations the Lorentz boost effects  in
 the wave function of a moving particle are always ignored.
 Probably this can be justified only at low  momentum transfer (cf. e.g. Fig. \ref{waves}).
 Contrarily, the BS amplitude, Eq.  (\ref{BSA}), is a covariant object by definition. Consequently,
 by applying the Lorentz transformation to the amplitude found in the
 rest system of the two-body system, one can obtain the BS solution in any system of reference.
However, mathematically such a procedure is quite
involved (see, e.g. Ref. \cite{static}), and in real calculations it becomes rather
difficult to separate and investigate pure boost and other relativistic effects.
In order to avoid such difficulties it is tempting to find
a covariant representation of the amplitude, i.e., a representation within which the
partial amplitudes are  frame independent.
Such a representation can be accomplished if one observes that, for instance in the deuteron case,
one can form a new covariant basis of $4\times 4$ matrices
from the available four independent matrix structures, $\hat \xi$, $(p\xi)\hat I$,
$S(p_1)^{-1}=\displaystyle\frac{\hat P}{2}+\hat p-m$, and
 $S(p_2)^{-1}=\displaystyle\frac{\hat P}{2}-\hat p-m$, where $\xi$ is the deuteron
  polarization four-vector.
Then the BS amplitude can be presented in the form
\begin{eqnarray}
\Psi_D(P,p) = \left [  { h_1}  \hat{\xi} +  { h_2}  \frac {(p \xi)}{m} \right] +
\frac {\hat{P}/2+\hat{p}-m}{m}\left [
 { h_3}  \hat{\xi} +  { h_4}  \frac {(p \xi)}{m}\right] +
\nonumber \\[0mm]
\left [  { h_5}
 \hat{\xi} +  { h_6}   \frac {(p \xi)}{m}\right ] \frac {\hat{P}/2-\hat{p}+m}{m}+
\frac {\hat{P}/2+\hat{p}-m}{m}
\left [  { h_7}
  \hat{\xi} + { h_8}   \frac {(p \xi)}{m}\right ] \frac {\hat{P}/2-\hat{p}+m}{m},
\label{covariant}
\end{eqnarray}
where  the ${ h_i}$ are the sought covariant partial amplitudes. Obviously, these new amplitudes can be
related to the non covariant ones by projecting (\ref{covariant}) onto the corresponding basis.
Explicit relations of covariant  amplitudes ${ h_i}$ in terms of spin-angular components can be found
in Ref. \cite{static}.
Below we present an illustration of how the covariant amplitudes (\ref{covariant})
can be used to analyze the process of backward elastic $pD$ scattering.
This process has been investigated in
the non relativistic Schr\"odinger approach within which the
    wave function in initial and final states has been
 taken  for the deuteron at rest. The corresponding
non relativistic  helicity amplitudes and the cross section can be written as~\cite{dp}
\begin{eqnarray}
&&
{\cal A}_{NR}\,= \,\left( u(P_{lab})+\frac{w(P_{lab})}{\sqrt{2}}\right) ^2 P_{lab}^2,\label{alad}, \quad
{\cal B}_{NR}\,= \,-\frac{3}{2}w(P_{lab})\,\left( 2\sqrt{2}u(P_{lab})-w(P_{lab}) \right ) P_{lab}^2,
\nonumber\label{blad}\\
 &&
{\cal C}_{NR}\,= \,\left( u(P_{lab})+\frac{w(P_{lab})}{\sqrt{2}}\right)
\left(u(P_{lab})-\sqrt{2}w(P_{lab})\right )P_{lab}^2,\label{clad},
 \label{dlad} \\
&&
{\cal D}_{NR}\,=\,  \frac{3w(P_{lab})}{\sqrt{2}}\left( u(P_{lab})+\frac{w(P_{lab})}{\sqrt{2}}\right)P_{lab}^2,\quad
\frac{d\sigma_{NR}}{d\Omega}\,=\,3\,\left( u^2(P_{lab})+ w^2(P_{lab})\right)^2 P_{lab}^4,
\label{crosrekal}
\end{eqnarray}
where $P_{lab}$ is the momentum of the nucleon in the laboratory system,
$u$ and $w$ are the two components of the deuteron wave function and the partial
helicity amplitudes are defined as
\begin{eqnarray}
{\cal F} = {\cal A}\,(\mbox{\boldmath{$\xi$}}_M\mbox{\boldmath{$\xi$}}^+_{M'})+
{\cal B}\,(\mbox{\boldmath{$n$}}\mbox{\boldmath{$\xi$}}_M)(\mbox{\boldmath{$n$}}\mbox{\boldmath{$\xi$}}^+_{M'})
+i{\cal C}\,(\mbox{\boldmath{$\sigma$}}\cdot [\mbox{\boldmath{$\xi$}}_M\times\mbox{\boldmath{$\xi$}}^+_{M'}])
+i{\cal D}\,(\mbox{\boldmath{$\sigma$}}\mbox{\boldmath{$n$}})(\mbox{\boldmath{$n$}}\cdot [\mbox{\boldmath{$\xi$}}_M\times\mbox{\boldmath{$\xi$}}^+_{M'}]).
\label{amplit}
\end{eqnarray}
In Eq. (\ref{amplit}) $\mbox{\boldmath{$\xi$}}_M$ is the polarization 3-vector of the deuteron at rest,
   ${\cal F}$ is the total amplitude of the process. This amplitude
can be found by calculating the corresponding one-nucleon-exchange Feynman diagram and  using
the covariant BS amplitudes (\ref{covariant}) which allow to  write
 explicitly the  dependence  on the polarization vector $\mbox{\boldmath{$\xi$}}_M$ and,
 consequently, to determine the
 partial amplitudes $\cal A,B,C$ and $\cal D$. The result of a calculation,
with preserving only the main $"++"$   components, is \cite{dp}
\begin{eqnarray}
\frac{d\sigma_0}{d\Omega} & = & 3\,\left( \Psi_S^2(P_{lab})
+ \Psi^2_D(P_{lab})\right)^2\,P_{lab}^4
\left( 1+\frac{P_{lab}^2}{2m^2}+\frac{29P_{lab}^4}{16m^4}+
\frac{83P_{lab}^6}{32m^6} +\cdots\right ),
\label{crospositive}
\\[2mm]
{\cal A}_0 & = &   P_{lab}^2\,\left(
\Psi_S(P_{lab})-\frac{\Psi_D(P_{lab})}{\sqrt{2}}\right) ^2
\,{\cal L}(P_{lab}),
\label{aBS}\\
{\cal B}_0 & = &  P_{lab}^2\
\frac{3}{2}\Psi_D(P_{lab})\,\left( 2\sqrt{2}\Psi_S(P_{lab})
+ \Psi_D(P_{lab})\right )
\,{\cal L}(P_{lab}),
\label{bBS}\\
{\cal C}_0\,& = &  P_{lab}^2\,\left( \Psi_S(P_{lab})
-\frac{\Psi_D(P_{lab})}{\sqrt{2}}\right)
\left(\Psi_S(P_{lab})+\sqrt{2}\Psi_D(P_{lab})\right )
\,{\cal L}(P_{lab}),
\label{cBS}\\
{\cal D}_0 & = & -\, P_{lab}^2\,
\frac{3}{\sqrt{2}}\Psi_D(P_{lab})
\left( \Psi_S(P_{lab})-\frac{\Psi_D(P_{lab})}{\sqrt{2}}\right)\,
{\cal L}(P_{lab}),
\label{dBS}
\end{eqnarray}
where $\Psi_{S,D}$ are the main, $"++"$,  components  of the deuteron
 (see Fig. \ref{waves}),
and   ${\cal L}(P_{lab})=
\left( 1+\frac{P_{lab}^2}{4m^2}+\frac{7P_{lab}^4}{8m^4}
+\cdots \right )$. It can be seen  that the analytical expressions
for relativistic and non relativistic helicity amplitudes are rather similar,
  except for the factor ${\cal L}(P_{lab})$ which
depends   on the velocity of the deuteron and obviously describes the Lorentz boost effects.
Other pure relativistic corrections appear in the cross section from the contribution of
the negative $P$-waves \cite{dp}.
In Fig. \ref{pD}, we exhibit results of calculations of the differential
cross section with emphasis   on different contributions. The notation in Fig. \ref{pD} is quite
self-explanatory. One can conclude  that at large values of the initial energy the relativistic
effects increase and become significant  at $P_{lab} \ge 0.35-0.4\ GeV/c$.

\subsection{The one-iteration approximation}\label{oneiter}
 As mentioned above, the  basis of Dirac matrices
  turns out  to be  the most appropriate representation
   in solving numerically the BS equation.
 Transitions to other bases can be accomplished
 by performing the corresponding unitary transformations.
 In such a way one can obtain numerical solutions of the BS equation
 in representations, which are most relevant for the considered problem
 and one can investigate numerically different aspects of the process.
 However  besides numerical investigations, it is useful  to have a mathematical
  method for analytical estimates of the considered effects.
Remind, that often
  in non relativistic approaches
 one takes into account relativistic corrections by  calculating
 additional diagrams with meson exchange currents, nucleon-antinucleon pair currents etc.
  It is intuitively clear that
some of these corrections must  be included already in the relativistic amplitude via negative
$P$ waves and Lorentz boost effects which are absent in the non relativistic calculations.
An estimate of the role of $P$ waves can be accomplished
by solving the BS equation in the so-called one-iteration approximation (OIA).
The spirit of the method follows from
the observation  that if, when solving numerically  the BS equation by iterations,
the trial function is
close to the exact solution, the latter one can be obtained by performing only
one or two iterations.

As illustrated in Fig. \ref{waves},  at low values of the relative momentum
$p$ the exact solution coincides, to a good accuracy, with the non relativistic
solution of the Schroedinger equation. This is a hint  that, using  the non relativistic deuteron
wave function as trial function,  the number
of iterations in solving the BS equation can be strongly reduced.
Moreover, at low and intermediate values of $p$ the first iteration with the
non relativistic solution as trial function  can serve as a
 good approximation of the exact solution.

\begin{figure}
\centering \includegraphics[width=0.5\textwidth]{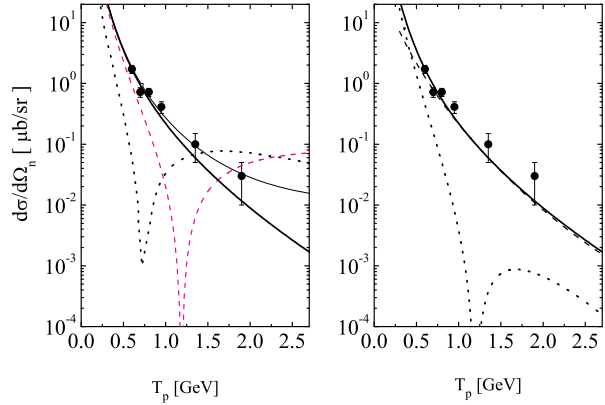}
\caption{The cross section for the reaction $pD\to (pp) n$.
Left panel: dotted and dashed lines represent the
 calculations within the nonrelativistic and relativistic impulse approximation, respectively;
the two solid
lines are the results which include the final state interaction in the (pp) - pair
in a  $^1S_0$ state calculated within the one-iteration approximation.
The  lower and upper solid lines have been obtained with and without cut-off form factors
in the BS equation.
Right panel: the relative contribution of the deuteron relativistic
 $S$ (dotted line) and $D$ (dashed line) components  to the total cross section.
 Experimental data  are from  \cite{exper}} \label{twofold}.
\end{figure}

 As an example, let us consider the BS equation for the
 deuteron partial vertices in the spin-angular harmonics basis (\ref{gammadecomp}):

\begin{eqnarray}
&& G_\alpha(k_0,|\bk|) = -\sum_bg_b^2\int \frac{\bp^2 d|\bp| \, d
p_0}{4\pi^2} W_{\alpha\beta}(k,p) \, \Psi_\beta(p_0,|\bp|),
\label{wab} \\
&& W_{\alpha\beta}(k,p)\equiv \sum_{lm}
\frac{Q_l(z)}{2\pi|\bp||\bk|}\int\! d\Omega_k\, d\Omega_p\,
Y_{lm}(\bp)\,Y^*_{lm}(\bk)\, \mbox{Tr}\! \left
[\Gamma_\alpha^+(-\bk)\gamma_b\Gamma_\beta(\bp) \gamma_b\right ].
\nonumber
\end{eqnarray}
The OIA method implies:
i)  in (\ref{wab})  all "negative" waves are put equal to zero,
ii)   the   $k_0$   and   $p_0$  dependencies in the partial kernels
$W_{\alpha\beta}(k,p)$ and in  vertices $G_{^3S_1^{++}}$ and  $G_{^3D_1^{++}}$ are disregard, i.e.
$   W_{\alpha\beta}( k, p)\simeq W_{\alpha\beta}(|\bk|,|\bp|), $
$   G_{++}(p_0,|{\bf p}|) \simeq G_{++}(0,|{\bf p}|)$,
iii) the trial function  is taken as solution  of the  non relativistic Schroedinger
equation for the deuteron, and
iv) the negative (P) waves are obtained    only  by  one iteration in Eq. (\ref{wab}).
For instance, for  the pseudo-scalar isovector exchange we get
\begin{eqnarray}
W_{P_3^{-+}\to S^{++}}=-W_{P_3^{+-}\to S^{++}}=\sqrt{2}W_{P_1^{+-}\to S^{++}}=\sqrt{2}W_{P_1^{-+}\to S^{++}}=
{\cal N} \sqrt{2}\left [ Q_0(y) |\vec k| - Q_1(y) |\vec p| \right ],\label{v61}\\
 W_{P_3^{+-}\to D^{++}}=-
W_{P_3^{-+}\to D^{++}}=\sqrt{2}W_{P_1^{+-}\to D^{++}}=
\sqrt{2}W_{P_1^{-+}\to D^{++}}=
{\cal N} \left [- Q_1(y) |\vec p| + Q_2(y) |\vec k| \right ],\label{v521}
\end{eqnarray}
where ${\cal N} = -\sqrt{3}/(2|\vec p||\vec k|E_k)$,
$E_k = \sqrt{{\bf k}^2+m^2}\sim m$,    $m$ being the nucleon mass.
 In  coordinate space one has
\begin{eqnarray}
&&
\int \frac{p^2dp}{2|\vec p||\vec k|}\left [ |\vec k | Q_0(z) -|\vec p| Q_1(z)\right ]
\Psi_S(|\vec p|) = \int\, dr \frac{\Psi_S(r)}{r}{\rm e}^{ -\mu r} (1+\mu r)j_1(kr)\label{ur},\\[2mm]
&&
\int \frac{p^2dp}{2|\vec p||\vec k|}\left [ |\vec p | Q_1(z) -|\vec k| Q_2(z)\right ]
\Psi_D(|\vec p|) =- \int\, dr \frac{\Psi_D(r)}{r}{\rm e}^{ -\mu r} (1+\mu r)j_1(kr)\label{wr},
\end{eqnarray}
where $\mu$ stands for the pion mass. Then  the result for the negative $P$-waves for
the pseudo-scalar one-boson exchange reads
\begin{equation}
\Psi_{P^{+-}_3(P^{+-}_1) }(P_{lab}  ) = -g_\pi^2
\frac{2\sqrt{3}}{M_dE_p'}
\int\limits_0^\infty dr \, \frac{{\rm e}^{-\mu r}}{r} \, (1+\mu r) \,
{\rm j_1}(r P_{lab})
\left [
N_u \, u(r) + N_w \, w(r)\right ],\label{uwr}
\end{equation}
where $u(r)$ and $w(r)$ are the non-relativistic deuteron wave functions
in the coordinate representation, and $g_\pi^2 \approx$ 14.5
is the pion-nucleon coupling constant.
The normalization factors are
$N_u = \sqrt{2}$ (1) and $N_w =$  -1 ($\sqrt{2}$) for
$P^{+-}_3$ $(P^{+-}_1)$ waves.
With this   negative-energy waves one may estimate
the origin of the relativistic corrections computed
within the non-relativistic limit as additional contribution to the impulse
approximation diagrams, such as meson exchange currents and
$N\bar N $ pair production currents. As an example
we present here the relativistic corrections to the partial amplitude
  $\cal A$ (\ref{amplit})
\begin{eqnarray} \hspace*{-1cm}
\delta {\cal A} &=& -g_\pi^2 P_{lab}^3
\frac{192 \sqrt{2}\pi}{M_dE_p'}
\int\limits_0^\infty dr \, \frac{{\rm e}^{-\mu r}}{r} \, (1+\mu r) \,
{\rm j_1}(r P_{lab})
\left [
\sqrt{2} \, u(r) - \, w(r)\right ]
\left(\Psi_S- \frac{1}{\sqrt{2}} \Psi_D\right )
\label{dp14}
\end{eqnarray}
which is similar to the expressions obtained in non-relativistic evaluations
of the so-called ``catastrophic'' and pair production   diagrams
in electro-disintegration processes of
the deuteron \cite{disintegration} and also similar to the
results of a computation of the triangle diagrams usually considered
in the elastic $pD$ processes \cite{dp,wilkin,nakamura}.
Another example
concerns the  calculation  of the effects of final state interaction
in reactions of deuteron break-up with two nucleons in the final state with
low relative energy  \cite{exper}.
Exact  calculations require also the
 solution of the BS equation for the scattering state in the continuum.
At the considered energies one can avoid the involved procedure
of solving the BS equation  in the continuum by employing  the OIA in the same manner
as described above. In Fig. \ref{twofold} we present results of calculations of the
cross section for the reaction $pD\to (pp) n$ within the OIA for the BS amplitudes
and a comparison with non relativistic results and available experimental data. It can be seen
 that relativistic calculations provide a  much better agreement with data.

\section{Exhausting method}
\label{exhaus} \noindent
Notice  that the   above mentioned methods are higly suitable for solving the BS
equation by iterations,   which provide the  solution of ground state of the system.
This is quite acceptable in most calculations  for the deuteron
and light mesons. However, heavier mesons (e.g. $J/\Psi$, $D$ mesons) possess an entire
mass spectrums   with various, experimentally known,  transitions to the ground state.
Consequently, one needs a generalization of the iteration procedure which would allow
 to find the energy spectrum of the excited states and the corresponding amplitudes.
 The "exhausting" (depletion) method \cite{nashYaph} is the one
 within which  the ground and
 excited states of the system can be found by solving the BS equation by iteration.
 The main idea of the method can be illustrated in the case of the scalar BS
equation with scalar interaction. The corresponding equation reads
\begin{eqnarray}
\varphi_{JM}(p)=i\,g^2\,
S(p_1)S(p_2)\int\!\frac{d^4p'}{(2\pi)^4}~\frac{1}{(p-p')^2-\mu^2+i\varepsilon}~
\varphi_{JM}(p'),\label{SBSEp}
\end{eqnarray}
where the scalar propagator is
$S(p)=1/(p^2-m^2+i\varepsilon)$ and  $JM$ are the total angular momentum and its
projection.
It is more convenient the introduce the BS equation for the
vertex function, $\varphi_{JM} (p)=S(p_1)~\CG_{JM}
(p)~S(p_2)$. Then in Euclidian space one has
\begin{eqnarray}
&&\CG_{JM} (p)=\lambda~\int d^4p'~
\underbrace{\frac{1}{(2\pi)^4}~\frac{1}{[(E^2_{\bp} +
p_0^2-\frac{1}{4}M_B^2 )^2 + p_0^2 M_B^2 ]
[(p-p')^2 +\mu^2 ]}}_{K(p,p';M_B)}~\CG_{JM}(p'),
\label{scal}
\end{eqnarray}
where $\lambda\equiv g^2$. Obviously, at fixed $M_B$ Eq. (\ref{scal})
is a Fredholm type  eigenvalue equation relative to
$\lambda_{iJ}=g^2_{iJ}$ with eigenfunctions
$\CG_{JM}^{i} (p)$. The set of eigenvalues
$\lambda_{iJ}\equiv\lambda_{iJ}(M_B)$ is also called the spectrum of
(\ref{scal}) relative to the coupling constant.
It is straightforward to establish  a correspondence between the
mass spectrum of the equation and its spectrum on coupling constant;
it is sufficient to  find the spectrum
 $\lambda_{iJ}(M_B)$ at all allowed values of
$M_B,~0\leq M_B\leq 2 m$ and to inverse the problem, i.e.
to fix  $\lambda$ at its actual value and to find the corresponding value of $M_B$.
In such a way, each $\lambda_i$ receives
its particular value of $M_B$, which forms the mass spectrum
of the equation. From the rigorous Hilbert-Schmidt theory of symmetric
integral equations  it is known that:
i) the spectrum of Eq. (\ref{scal}) is discrete and real
with a nondecreasing sequence of $\lambda_{iJ}$
$|\lambda_{1}|\leq|\lambda_{2}|
 \leq...\leq|\lambda_{n}|\ldots$,
ii) the corresponding eigenfunctions  $\CG^i_{JM}(p)$ and the corresponding
amplitudes $\varphi^i_{JM}(p)$ form a bi-orthogonal basis
$
\int
d^4p~\varphi^{i*}_{JM}(p)~\CG^j_{J'M'}(p)=\delta_{ij}\delta_{JJ'}
\delta_{MM'}$
 within which the kernel  $K(p,p';M_B)$ can be represented as
\begin{eqnarray}
K(p,p';M_B)=\sum\limits_{
i=1}^{\infty}\sum\limits_{JM}\frac{1}{\lambda_{iJ}}~
\CG^i_{JM}(p)~\varphi^{i*}_{JM}(p').\label{gilb}
\end{eqnarray}
For the scalar equation the partial amplitudes have a  particularly simple form,
$\CG^i_{JM}(p)=\frac{1}{|{\bp}|}G^i_{J}(\tilde p)
{\mbox{\large {Y}}}_{JM}(\Omega_p),\label{verpar}
$
where for brevity  the notation  $\tilde
p=(p_0,~|{\bp}|)$ has been introduced.  Then the partial vertices
$G^i_J(\tilde p)$ read
\begin{eqnarray}
&&
G^i_J(\tilde p)=\lambda_{iJ}~
\int d\tilde p'~K^J(\tilde p;\tilde p';M_B)~
G^i_J (\tilde p'),\label{partsial}\\
&&K^J(\tilde p;\tilde p';M_B)=
\frac{1}{(2\pi)^3}~\frac{1}{[(E^2_{\bp} +
p_0^2-\frac{1}{4}M_B^2 )^2 + p_0^2 M_B^2 ]}
~Q_J({ y}). \label{ktildp1}
\end{eqnarray}
 Notice  that   the partial vertices $G^i_J(\tilde p)$
 together with the partial amplitudes $\varphi_J^i$ also form
 a bi-orthogonal basis, hence the partial kernels
$K^J(\tilde p;\tilde p';M_B)$ can also be written in the form
\begin{eqnarray}
K^J(\tilde p;\tilde p';M_B)=
\sum\limits_{i=1}^{\infty}\frac{1}{\lambda_{iJ}}~
G^i_{J}(\tilde p)~\varphi^i_{J}(\tilde p').\label{gilbpar}
\end{eqnarray}
Now we are in the position to formulate the   exhausting method
for the Eq. (\ref{scal}).  Consider the problem of finding the lowest eigenvalue
$\lambda_{1J}$ at fixed  $J$ and $M_B$. Schematically, Eq.
(\ref{partsial}) read
\begin{eqnarray}
\frac{1}{\lambda_{1J}}~G^i_J=K^J~G^i_J.\label{prtsbr}
\end{eqnarray}
Let now choose a trial function $\chi(\tilde p)$ and form a
sequence of functions  $K^J~\chi$, $(K^J)^2\chi$, ...,
$(K^J)^n\chi$. The  trial function $\chi$
can be decomposed on the complete set of functions $G^i_J$,
$ 
\chi=\sum\limits_{i=1}^{\infty}c_i~ G^i_J,
$ 
where
$c_i=\int d\tilde p~\varphi^i_{J}(\tilde p)~
\chi(\tilde p)$ and from Eq. (\ref{prtsbr})  one gets

\noindent
\begin{eqnarray}
&&K^J~\chi=\sum\limits_{i=1}^{\infty}c_i~ \frac{1}{\lambda_{iJ}}~ G^i_J,
\quad
(K^J)^2\chi=\sum\limits_{i=1}^{\infty}c_i~ \frac{1}{\lambda_{1J}^2}~G^i_J,\ldots,\quad
(K^J)^n\chi=\sum\limits_{i=1}^{\infty}c_i~ \frac{1}{\lambda_{1J}^n}~ G^i_J.
\label{posled}
\end{eqnarray}

\noindent
At each $n$-th iteration one considers the ratio
$(K^J)^n\chi/(K^J)^{n-1}\chi$, which at large enough $n$ provides the lowest eigenvalue
of the spectrum:
\begin{eqnarray}
\frac{(K^J)^n\chi}{(K^J)^{n-1}\chi}=\frac{\sum\limits_{i=1}^{\infty}c_i~
\frac{1}{\lambda_{iJ}^n}~ G^i_J}
{\sum\limits_{i=1}^{\infty}c_i~\frac{1}{\lambda_{iJ}^{n-1}}~ G^i_J}=
\frac{1}{\lambda_{1J}}~\frac{c_1~G^1_J+\sum\limits_{i=2}^{\infty}c_i~
\left(\frac{\lambda_{1J}}{\lambda_{iJ}}\right)^n~ G^i_J}
{c_1~G^1_J+\sum\limits_{i=2}^{\infty}c_i~
\left(\frac{\lambda_{1J}}{\lambda_{iJ}}\right)^{n-1}~ G^i_J}
\stackrel{n\to \infty}{\Longrightarrow}
\frac{1}{\lambda_{1J}}.\label{vot}
\end{eqnarray}

\begin{figure}[h]
\centerline{\includegraphics[width=0.43\textwidth]{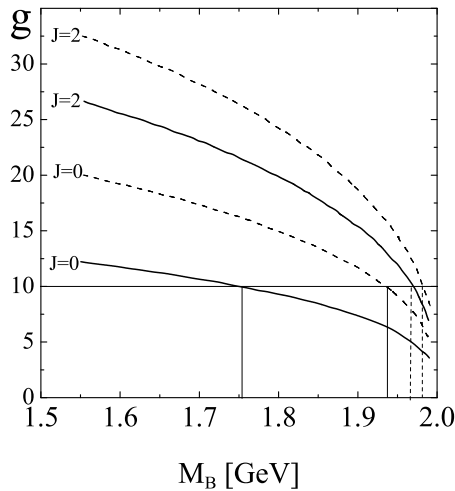}}
\vskip -0.3cm \caption{The coupling constant $g$ vs. the value of the mass   $M_B$ of the bound system for two
values of the total angular momentum $J=0,2$. The solid curves correspond to the ground state, while
the dashed lines are for the first excited state. Intersections of the horizontal line  $g=10$
with the respective curve determine the energy spectrum of the system: $M_B\simeq 1.75\ GeV$,
 $M_B\simeq 1.94\ GeV$ for $J=0$ and $M_B\simeq 1.97\ GeV$, $M_B\simeq 1.98\ GeV$ for $J=2$.}
\label{mass}
\end{figure}
Equations (\ref{posled}) and (\ref{vot}) demonstrate how one finds the ground state solution, i.e. the
eigenvalue $\lambda_{1J}$ and the eigenfunction $G_{1J}$ can be found. Now, to find the next  excited
eigenvalue $\lambda_{2J}$, one defines a new kernel $K^{J,1}$ for which the new BS equation
 has the same spectrum $\lambda_{iJ}$ and eigenfunctions $G_{iJ}$ ($i=2...n$)
 except the first state $i=1$. For such a kernel, $\lambda_{2J}$ and $G_{2J}$
 would serve as ground state solution. The new kernel which obeys these conditions is

\begin{eqnarray}
K^{J,1}(\tilde p,\tilde p';M_B)=K^J(\tilde p,\tilde p';M_B)-
\frac{1}{\lambda_{1J}}~G^1_{J}(\tilde p)~\varphi^1_{J}(\tilde p')=
\sum\limits_{i=2}^{\infty}\frac{1}{\lambda_{iJ}}~
G^i_{J}(\tilde p)~\varphi^i_{J}(\tilde p').\label{glb1}
\end{eqnarray}
It can be seen that all  eigenfunctions $G^i_J$ of the kernel
$K^J$, except $G^1_J$, are the eigenfunctions of the new kernel
$K^{J,1}$ with the same eigenvalue spectrum $\lambda_{iJ}$.
Obviously, the lowest eigenvalue of
$K^{J,1}$ is $\lambda_{2J}$, which can be found by the same iteration procedure, i.e. via
 Eqs. (\ref{posled}) and (\ref{vot}).
Analogously one finds the next solutions $i=3,4\ldots$ from the recurrent  kernel

\begin{eqnarray}
K^{J,n}(\tilde p,\tilde p';M_B)=K^{J,n-1}(\tilde p,\tilde p';M_B)-
\frac{1}{\lambda_{nJ}}~G^n_{J}(\tilde p)~\varphi^n_{J}(\tilde p')=
\sum\limits_{i=n+1}^{\infty}\frac{1}{\lambda_{iJ}}~
G^i_{J}(\tilde p)~\varphi^i_{J}(\tilde p').\label{glbn}
\end{eqnarray}

As an example of application of the method,
 results of calculations of the  eigenvalues $g=\frac{1}{\lambda}$
as functions of the binding mass $M_B$ are presented in Fig. \ref{mass}
for first two ground states of a system of two equal masses particles
with total angular momentum $J=0$ and $J=2$  (solid lines)
and two excited states (dashed lines), respectively. It is clear  that, if the
coupling constant is known from independent experiments,
 the mass spectrum of the system can be easily  recovered
(see illustration in  Fig. \ref{mass} for $g=10$).

A generalization of the method for the spinor-spinor BS equation
can be found in \cite{nashYaph}.

\section{Hyperspherical harmonics}\label{hyp}
\begin{figure}
\vskip 3mm
\centerline{\includegraphics[width=0.55\textwidth]{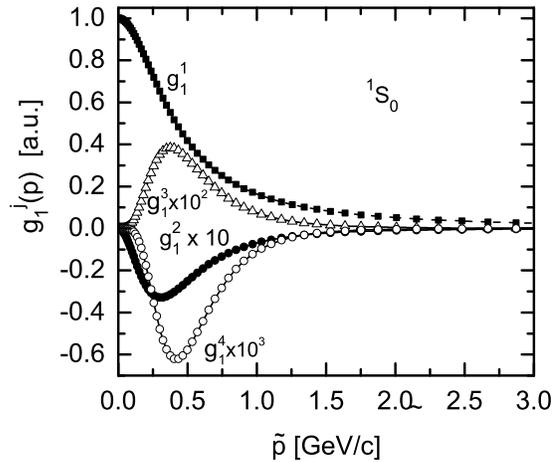}}
\vskip -0.3cm \caption{Functions $g_1^j, j=1,\ldots,4$, eq.  (\ref{s0p1})-(\ref{s0p2}).
Full squares correspond to
$g_1^1$;  full circles to $g_1^2$ multiplied by 10; triangles
to $g_1^3$ multiplied by 100; open circles to $g_1^4$ multiplied by
1000; the solid lines correspond to the  fitted functions  $g_1^j
$
 by formula (\ref{fit}).
The overall   normalization constant is arbitrary. }
\label{hyperF}
\end{figure}

The considered methods are convenient for solving numerically the BS equation
by an iteration procedure. The obtained solution presents a set of two-dimensional
arrays of the partial amplitudes $\psi_i$, cf. Eq. (\ref{amp6}). In practice, such a representation
of the solution can cause difficulties in specific calculations  which require two-dimensional
interpolations of such arrays, e.g. in attempts to determine analytical
parametrizations  of the numerical solution, in performing the inverse Wick rotation, etc.
This is particularly awkward to use such solutions if one solves the BS equation for
$q\bar q$ bound systems (mesons) in nonperturbative QCD.  In the later case, the BS equation
must be solved simultaneously   with the Dyson-Schwinger equation to generate dynamically  the masses
of the constituent quarks. Numerical
solutions in  form of one-dimensional arrays would be
much more appropriate for tthese problems.
 This can be achieved by decomposing the partial amplitudes over
 the three-dimensional hyperspherical harmonics basis  in Euclidian space

\begin{eqnarray}
Z_{nlm}(\chi,\theta,\phi)=X_{nl} (\chi) Y_{lm}(\theta,\phi); \quad
X_{nl}(\chi)=\sqrt{\frac{2^{2l+1}}{\pi} \frac
{(n+1)(n-l)!l!^2}{(n+l+1)!}} \sin^l\chi C_{n-l}^{l+1}(\cos \chi),
\label{hyper}
\end{eqnarray}
where $C_{n-l}^{l+1}$    are
the usual Gegenbauer polynomials,
$\cos \chi =\displaystyle\frac{p_0}{\hat p};\,\,
\sin\chi =\displaystyle\frac{|{\bf p}|}{\hat p};\,\,
\hat p=\sqrt{p_0^2+{\bf p}^2}
$.
The BS vertices (amplitudes) are decomposed over the complete set
of spin-angular harmonics.
\begin{eqnarray}
{\cal G}(p_0,\bp)&=&\sum\limits_\alpha g_\alpha(p_0,|\bp|)
\,\Gamma_\alpha(\bp) \label{ng}
\end{eqnarray}
with the coefficients $g_\alpha(p_0,|\bp|)$ as
\begin{eqnarray}
g_\alpha(p_0,\bp)=\sum_{j,n,m}^\infty g^j_\alpha(\hat p)\ Z_{jlm} (\chi_p,\theta_p,\phi_p).
\label{po}
\end{eqnarray}
Note  that, in order to be able to carry out the $\chi_p,\theta_p,\phi_p$ - integrations
analytically, the spin-orbital basis must be slightly redefined
 (see Ref. \cite{fbBayer} for details).
The scalar part of the interaction kernel   is also represented in terms of   hyperspherical
harmonics
\begin{eqnarray}
\frac {1}{(p-k)^2+\mu^2}&=&2\pi^2 \sum_{nlm} \frac{1}{n+1}
V_n(\hat p,\hat k)Z_{nlm}(\chi_p, \theta_p,\phi_p)
Z_{nlm}^{*}(\chi_k, \theta_k,\phi_k),\nonumber
\\
V_n(\hat p,\hat k)&=& \frac {4}{(\Lambda_+ +\Lambda_-)^2}
\left( \frac {\Lambda_+ -\Lambda_-}{\Lambda_+
+\Lambda_-}\right)^n;\quad
\Lambda_\pm=\sqrt{(\hat p \pm \hat k)^2+\mu^2}. \label{ker}
\end{eqnarray}
By  inserting Eqs. (\ref{ng})-(\ref{ker}) into the BS equation one obtains a system
of one-dimensional integral equations for the partial amplitudes

\begin{eqnarray} &&
g_{1}(p_0,|\bp|)=\sum_{j=1}^\infty
  g_{1}^j(\hat p) \,X_{2j-2,0}(\chi_p);\quad
  g_{2}(p_0,|\bp|)=\sum_{j=1}^\infty
  g_{2}^j(\hat p) \,X_{2j-2,0}(\chi_p);
\label{s0p1}\\ &&
g_{3}(p_0,|\bp|)=\sum_{j=1}^\infty
 g_{3}^j(\hat p)  \,X_{2j-1,1}(\chi_p);\quad
g_{4}(p_0,|\bp|)=\sum_{j=1}^\infty
  g_{4}^j(\hat p) \,X_{2j,1}(\chi_p),
\label{s0p2}
\end{eqnarray}
where

\begin{eqnarray}
g^j_{1,2}(\hat p)&=&-g^2\ b_{1,2}\int\limits_{0}^\infty
\frac{d\hat k\, {\hat k}^3}{8\pi^2(2j-1)}\,V_{2j-2}(\hat p,\hat k)
\sum\limits_{n=1}^4\sum\limits_{m=1}^\infty
A^{1,2\ n}_{jm}(\hat k)\,g_n^m(\hat k),
\label{set1}\\
g^j_3(\hat p)&=&-g^2b_3 \int\limits_{0}^\infty\frac{d\hat k\,
{\hat k}^3}{8\pi^2\, 2j}\,V_{2j-1}(\hat p,\hat k)
\sum\limits_{n=1}^4\sum\limits_{m=1}^\infty
A^{3\ n}_{jm}(\hat k)\,g_n^m(\hat k),
\\
g^j_4(\hat p)&=&-g^2 b_4\int\limits_{0}^\infty\frac{d\hat k\,
{\hat k}^3}{8\pi^2(2j+1)}\,V_{2j}(\hat p,\hat k)
\sum\limits_{n=1}^4\sum\limits_{m=1}^\infty
A^{4\ n}_{jm}(\hat k)\,g_n^m(\hat k).\label{set4},
\end{eqnarray}
where $b_i$ and $A^{\alpha\ n}_{jm}$ are known coefficients, whose explicit expressions
 can be found in Ref.~\cite{fbBayer}.
It can be seen from Eqs. (\ref{set1})-(\ref{set4}) that the partial amplitudes within the
hyperspherical basis are functions of only one variable. Instead, the obtained
system of integral equations is infinite. At first glance, this may
cause unresolvable problems. However, in practice, when solving
the system (\ref{set1})-(\ref{set1})
a good convergence  can be achieved by considering only the few first terms
in (\ref{s0p1}) and  (\ref{s0p2}),
usually up to $j=3-4$ \cite{fbBayer}. In Fig. \ref{hyperF} we present an example of
the use of hyperspherical basis in solving the BS equation for the $^1S_0$ channel.
The obtained solution can be fitted by a simple formula
\begin{eqnarray}
\label{fit} g_1^j(\hat p)\simeq
 \left[\frac{\hat p^2}{\hat p^2+b_j^2}\right]^{j-1}
\sum\limits_{l=1}^4\frac{a_{jl}\,\hat p^{2l-2}}{(\hat p^2+b_j^2)^l}
\end{eqnarray}
which is extremely useful in further analyses  of the solution in Minkowski space. The fitted parameters
$a_j$ and $b_j$ can be found in Ref. \cite{fbBayer}.
\section{Summary}
 In this mini-review we briefly  considered  different methods to solve the BS equation in
 Euclidian space. It is shown that  the complete set of Dirac matrices is the most
 convenient way to decompose the BS amplitude and to find the partial amplitudes by iteration
 procedure. The transition to other representation, e.g. to the spin-angular harmonics basis
 or to the covariant representation can be accomplished by performing
 the corresponding unitary transformation. Examples of numerical solution for the
 deuteron and  applications for the calculation of matrix elements of observables  are presented,
 which illustrate    the role of relativistic corrections and Lorentz boost
 effects.

 A generalization of the iteration method to  find the energy spectrum of the
 BS equation is also discussed. It is shown that the Hilbert-Schmidt bi-orthogonal
 basis can define  a mathematical method, known as the "exhausting method", to find
 numerically the energy spectrum and the corresponding partial amplitudes for the excited
 states of two-body systems.

 The hyperspherical harmonics basis is shown to be the most appropriate one
 in finding the BS solution in form of one-dimensional numerical arrays. This is extremely
 convenient in analytical continuation of the solution back to Minkowski space
 and in solving simultaneously the Dyson-Schwinger and BS equation for the $q\bar q$ bound
 states (mesons) in nonperturbative QCD.

\end{document}